\documentclass[aps,twocolumn,showpacs,prl,preprintnumbers,amsmath,amssymb,superscriptaddress]{revtex4-2}
\usepackage{epsf}
\usepackage{graphicx}
\usepackage{sidecap}
 \usepackage{soul}
 \usepackage{array}
 \usepackage{amsmath}
 \usepackage{amssymb}
 \usepackage{dcolumn}
 \usepackage{epstopdf}
 \usepackage{bm}
 \usepackage{amsmath}
 
\usepackage{gensymb}
\usepackage{color}
\usepackage{hyperref}
\usepackage{soul}
\sethlcolor{green}
\usepackage{bbding}
\usepackage{sidecap}
\usepackage{float}

\hypersetup{
    colorlinks=true,
    citecolor=red,
    linkcolor=red,
    filecolor=blue,   
    urlcolor=blue,
}
 
\def \beq {\begin{equation}}
\def \eeq {\end{equation}}
\def\bibsection{\refname}
\renewcommand{\refname}{\noindent\textbf{References}}
\begin{document}
\title{Electronic structure of a layered altermagnetic compound CoNb$_4$Se$_8$}
\author{Anup Pradhan Sakhya} \thanks{These authors contributed equally.} \affiliation {Department of Physics, University of Central Florida, Orlando, Florida 32816, USA} 
\author{Mazharul Islam Mondal} \thanks{These authors contributed equally.}\affiliation{Department of Physics, University of Central Florida, Orlando, Florida 32816, USA} 
\author{Milo Sprague} \thanks{These authors contributed equally.} \affiliation{Department of Physics, University of Central Florida, Orlando, Florida 32816, USA} 
\author{Resham Babu Regmi} \affiliation{Department of Physics and Astronomy, University of Notre Dame, Notre Dame, IN 46556, USA}  
\affiliation{Stravropoulos Center for Complex Quantum Matter, University of Notre Dame, Notre Dame, IN 46556, USA}
\author{Arun K Kumay} \affiliation {Department of Physics, University of Central Florida, Orlando, Florida 32816, USA}
\author{Himanshu Sheokand} \affiliation{Department of Physics, University of Central Florida, Orlando, Florida 32816, USA}
\author{Igor. I. Mazin} \affiliation{Department of Physics and Astronomy, George Mason University, Fairfax, VA 22030, USA}
\author{Nirmal J. Ghimire} \affiliation{Department of Physics and Astronomy, University of Notre Dame, Notre Dame, IN 46556, USA}
\affiliation{Stravropoulos Center for Complex Quantum Matter, University of Notre Dame, Notre Dame, IN 46556, USA}
\author{Madhab Neupane} \thanks{Corresponding author:\href{mailto:madhab.neupane@ucf.edu}{madhab.neupane@ucf.edu}}\affiliation{Department of Physics, University of Central Florida, Orlando, Florida 32816, USA}

\date{\today}

\begin{abstract}
Recently, there has been a growing interest in altermagnetism, a novel form of magnetism, characterized by unique spin-splitting even in the absence of both net magnetic moments and spin-orbit coupling. Despite numerous theoretical predictions, experimental evidence of such spin-splitting in real materials remains limited. In this study, we use angle-resolved photoemission spectroscopy (ARPES) combined with density functional theory (DFT) calculations to investigate the electronic band structure of the altermagnet candidate CoNb$_4$Se$_8$. This material features an ordered sublattice of intercalated Co atoms within NbSe$_2$ layers. Magnetization and electrical resistivity measurements reveal the onset of antiferromagnetism below 168 K. Temperature-dependent ARPES data, supported by DFT calculations, uncover spin-split bands along the $\overline{\text{M}}$--$\overline{\Gamma}$--$\overline{\text{M}}$ high-symmetry direction. The observation of spin-splitting in this high-temperature altermagnet opens new avenues for exploring its electronic properties and potential applications in spintronic technologies. 
\end{abstract}

\maketitle
\section{I. Introduction}
\indent Traditionally, two primary categories of crystals with collinear magnetic order have been recognized: ferromagnetic and antiferromagnetic \cite{Neel1, Neel2}. Ferromagnetic and antiferromagnetic materials, distinguished by the alignment of magnetic moments within the material, exhibit distinct electronic properties and have long been established concepts in the field of magnetism. In ferromagnetic systems, the exchange interaction generates substantial magnetization and spin polarization in the electronic bands, breaking time-reversal ($\tau$) symmetry \cite{Nagaosa, Bader}. In contrast, compensated antiferromagnets maintain a global non-symmorphic time-reversal symmetry through translation or inversion symmetries relating opposite-spin sites and their electronic bands are typically spin-degenerate \cite{Helena}. Recent advancements in magnetism have introduced a novel type of magnetic order that exhibits characteristics of both antiferromagnets and ferromagnets. This emerging phenomenon, known as altermagnetism, is observed in materials termed altermagnets, which exhibit compensated net magnetization \cite{Smejkal, Libor}. 

\indent Similar to conventional collinear antiferromagnets, altermagnets break $\tau$ symmetry but maintain zero net magnetization. This occurs because the time-reversal operation is effectively nullified by a lattice rotation in altermagnets. Unlike antiferromagnets, altermagnets exhibit spin-polarized electronic bands and macroscopic responses akin to ferromagnets \cite{Libor, Mazin, Mazin1, Mazin2, Mazin3, Mazin4, Mazin5}. This spin-splitting of the electronic bands can be exceptionally large, as the mechanism is caused by different local crystal electric field environments at separate magnetic lattice sites connected by crystal rotational symmetries, independent of relativistic spin-orbit interactions \cite{Libor, Smejkal, Osu}. This has ignited significant research interest in altermagnetism and altermagnetic materials \cite{Mazin, Cui, Brekke, Ouassou}. This behavior requires the breaking of combined $P\tau$-symmetry, where \textit{P} represents parity and $\tau$ represents time-reversal symmetry operation \cite{Hodt}. Altermagnetism is anticipated to be prevalent in nature, manifesting in both three-dimensional and two-dimensional crystals across a wide range of structural and chemical compositions \cite{Libor}.

\indent This discovery of uniquely spin-polarized electronic bands provides new insights into the understanding and potential applications of magnetic quantum materials. Already several candidate altermagnets have been identified and their properties are the subject of extensive, albeit early, investigation. The experimental observation of spin-current \cite{Bose, BaiSpinTorqueRuO2, KarubeSpinTorqueRuO2, NakaOrganic, GonzalezHernandez, NakaPerovskite, MaPiezo, SmejkalMR} and the anomalous Hall effect \cite{FengAHERuO2, SmejkalAHE, BetancourtAHE} in compensated antiferromagnets was an early indication of altermagnetic phenomenon. Nevertheless, despite substantial theoretical progress only a few altermagnetic materials have been experimentally verified to exhibit band splitting using angle-resolved photoemission spectroscopy (ARPES) \cite{Osu}. The field is still in its early stages, with most of the focus on theoretical predictions and a limited number of experimental studies confirming the phenomenon. Among the proposed candidate materials, MnTe and CrSb have been experimentally studied \cite{BetancourtAHE, J, Lee, Osu, Rei}. 

\begin{figure*} 
	\includegraphics[width=17cm]{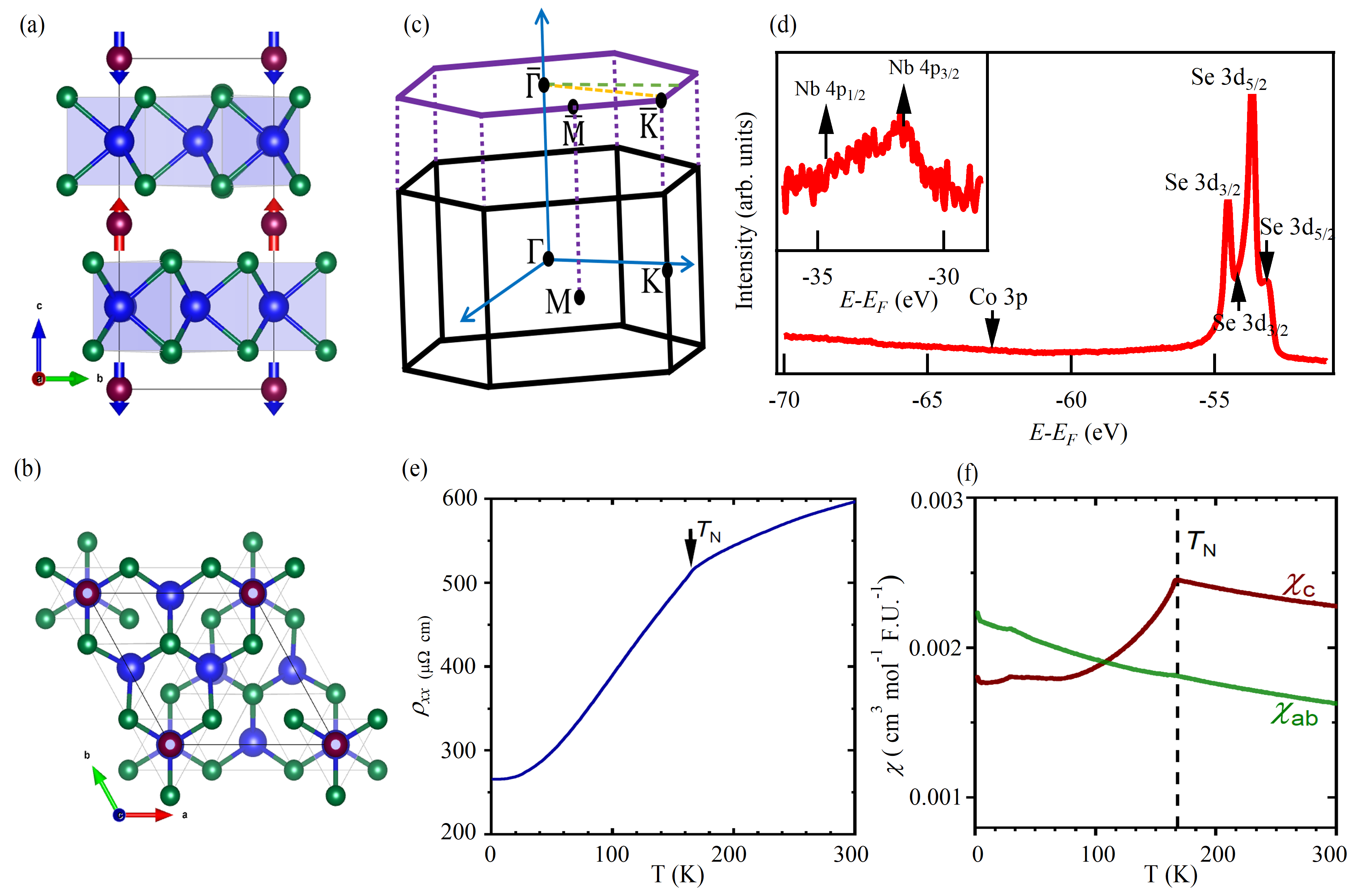} 
    \vspace{-1ex}
	\caption{Crystal structure and sample characterization. (a) Crystal structure of CoNb$_4$Se$_8$, with brown, blue, and green spheres representing Co, Nb, and Se atoms, respectively. (b) A view of the structure perpendicular to the \textit{a-b} plane. (c) Bulk Brillouin zone (BZ) and its projection onto the (001) surface BZ; high-symmetry points are marked. (d) Photoemission core-level spectrum confirming the high quality of CoNb$_4$Se$_8$ single crystals, with clearly identifiable Co, Nb, and Se core levels. (e) Temperature-dependent zero-field electrical resistivity. (f) Magnetic susceptibility ($\chi$) versus temperature with the field applied along the \textit{c}-axis, ($\chi_c$) and in the \textit{ab}-plane ($\chi_{ab}$).}
\label{fig2}
\end{figure*}
\indent Transition metal dichalcogenides (TMDs) are a fascinating family of materials known for their exotic properties, such as superconductivity, charge density waves (CDW), and non-trivial topology \cite{YLi, Ritschel, Sipos, Yang}. Due to their layered structure, various atoms can be intercalated into the van der Waals gaps of these TMDs. Recently, altermagnetism in CoNb$_4$Se$_8$ was identified through magnetometry, transport measurements, neutron scattering experiments, and density functional theory (DFT) calculations \cite{GhimireCoNb4Se8}. In this communication, we present a combined study of ARPES and DFT calculations to investigate the electronic structure of the layered intercalated transition metal dichalcogenide (ITMD) compound CoNb$_4$Se$_8$. Measurements of magnetization and electrical resistivity demonstrate the emergence of antiferromagnetic ordering at temperatures below 168 K. The Fermi surface topology and band structures along high-symmetry directions show strong agreement between ARPES data and DFT calculations. Notably, we observe band splitting along the $\overline{\text{M}}$--$\overline{\Gamma}$--$\overline{\text{M}}$ high-symmetry line in both ARPES measurements and DFT calculations. Our combined experimental and theoretical analysis confirms that CoNb$_4$Se$_8$ is an altermagnetic compound. With its high-temperature magnetism, CoNb$_4$Se$_8$ holds significant promise for exploring novel electronic and magnetic functionalities, potentially advancing future spintronic applications.

\begin{figure} 
	\includegraphics[width=8cm]{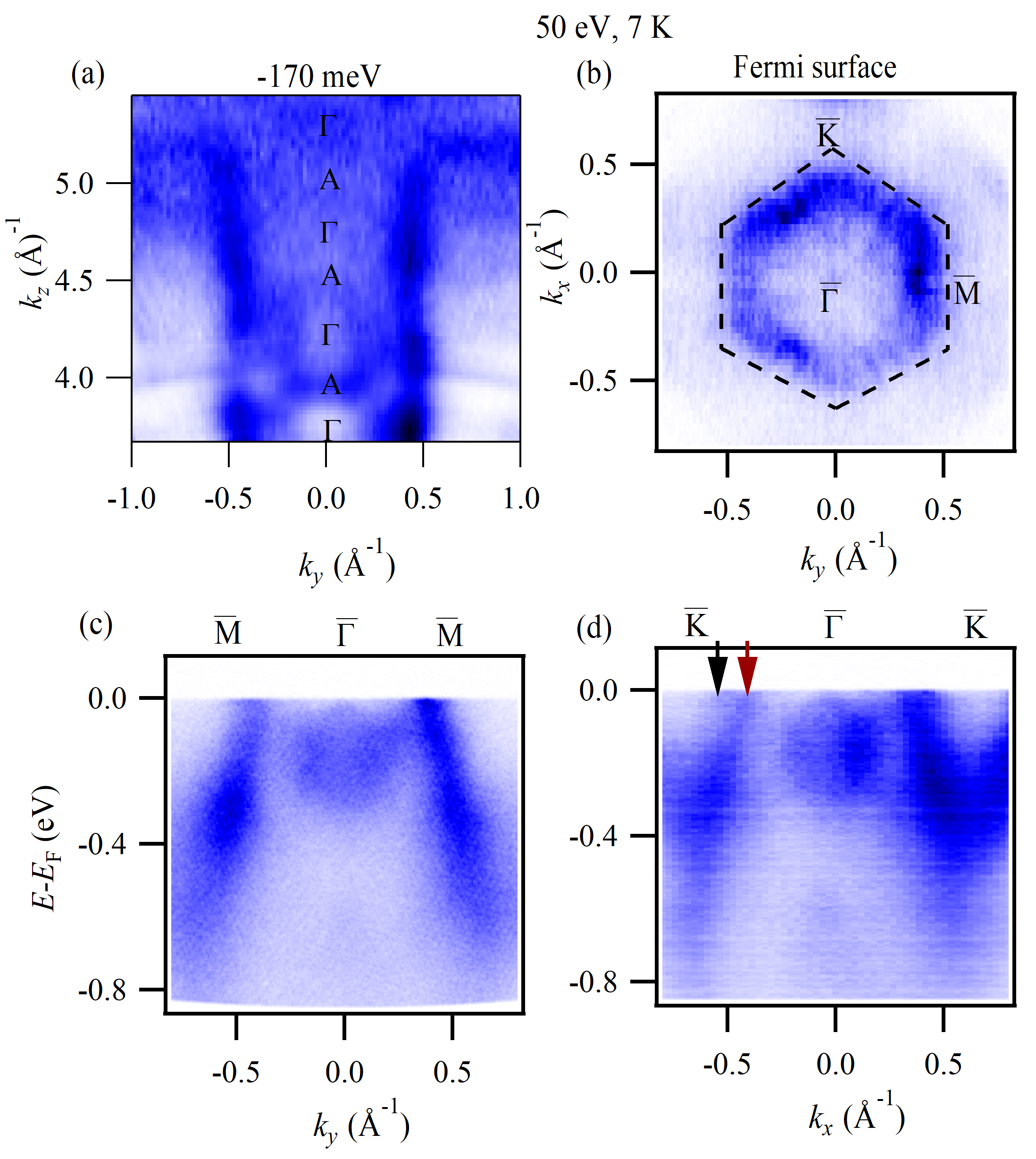} 
    \vspace{-1ex}
	\caption{Fermi surface and band dispersion along various high-symmetry directions. (a) \textit{k$_z$} dependent measurements along the $\overline{\text{M}}$--$\overline{\Gamma}$--$\overline{\text{M}}$ high-symmetry directions at the binding energy of -170 meV. High-symmetry points are marked in the plot. (b) ARPES measured Fermi surface of CoNb$_4$Se$_8$ at a photon energy of 50 eV and at a temperature of 7 K. High-symmetry points are indicated, with the Brillouin zone outlined by a dotted line. (c) Experimental band dispersion along the $\overline{\text{M}}$--$\overline{\Gamma}$--$\overline{\text{M}}$ high-symmetry line and (d) along the $\overline{\text{K}}$--$\overline{\Gamma}$--$\overline{\text{K}}$ high-symmetry line.}
\label{fig2}
\end{figure}

\section{II. Methods}
\noindent\textbf{Single crystal growth and characterization:}
\indent Single crystals of CoNb$_4$Se$_8$ were grown using chemical vapor transport with iodine as the transport agent. The crystal structure and chemical composition were confirmed through single crystal X-ray diffraction and energy-dispersive X-ray analysis. Further details on the synthesis method and crystal characterization can be found in Ref. \cite{GhimireCoNb4Se8}.\\

\noindent\textbf{ARPES measurements:}
\indent ARPES measurements were conducted at the Stanford Synchrotron Radiation Lightsource (SSRL), endstation 5-2. The measurements were performed at temperatures of $7$~K and $200$~K, corresponding to the altermagnetic and paramagnetic regimes, respectively. The pressure in the UHV chamber was maintained below $1\times10^{-10}$ Torr. The angular and energy resolutions were set to better than 0.2\degree and 15 meV, respectively. Measurements were performed using photon energies ranging from $40$~eV-$100$~eV with linear horizontal (LH) polarization.\\

\noindent\textbf{DFT calculations:}
\indent DFT calculations were performed using the projector augmented wave method as implemented in the Vienna ab initio simulation package (VASP) \cite{Kresse}, both for structural optimization and for exploring the magnetic ordering. For the latter, a doubled supercell consistent with a hypothetical in-plane propagation vector (1/2,0,0), similar to that in CoNb$_3$S$_6$, was employed. The results were fitted to a two nearest-neighbors Heisenberg Hamiltonian, yielding an excellent fit. The intraplanar exchange interaction was found to be ferromagnetic, J$\parallel$ $\sim$ 5.8 meV, while the interplanar exchange was antiferromagnetic, J $\perp$ $\sim$ 25.6 meV, both normalized to a unit magnetic moment. A generalized gradient approximation for the exchange and correlation functional \cite{Perdew} was used in all cases, without applying LDA+U or any other corrections beyond DFT. A \textit{k}-point mesh of up to 11$\times$11$\times$6 \textit{k}-point mesh (64 irreducible points) was used for structural optimization, and 48$\times$48$\times$25 for Fermi surface analyses.  The optimized structure was subsequently used in the augmented plane wave Wien2k code \cite{Blaha} for Fermi surface analysis. Schematics of both the crystal and magnetic structures were constructed using the three dimensional visualization software VESTA \cite{Momma}.

\section{III. Results and discussion} 
\indent The crystal structure of CoNb$_4$Se$_8$ shown in Figure 1(a,b), belongs to the hexagonal P6$_3$/mmc (No 194) space group and consists of offset 1H-NbSe$_2$ bilayers intercalated with Co atoms. In this structure, the bond center between two Co atoms along the \textit{c}-axis lacks inversion symmetry, influenced by the NbSe$_2$ prismatic layer, as shown in Fig. 1(a). This structure exhibits A-type antiferromagnetic ordering (shown by red and blue vectors in Fig. 1a), where Co planes are ferromagnetically aligned but stacked antiferromagnetically along the c-axis \cite{GhimireCoNb4Se8, Mandujano}. Such an arrangement meets the conditions for altermagnetism \cite{Mazin6, Mazin4, GhimireCoNb4Se8}. The Co atoms occupy 1/4 of the octahedral sites, forming an ordered centrosymmetric structure with a 2 $\times$ 2 superlattice based on the NbSe$_2$ unit cell. These Co atoms reside at the 2a Wyckoff position, causing the intercalation to double the in-plane lattice vectors.

\begin{figure*}
	\includegraphics[width=17cm]{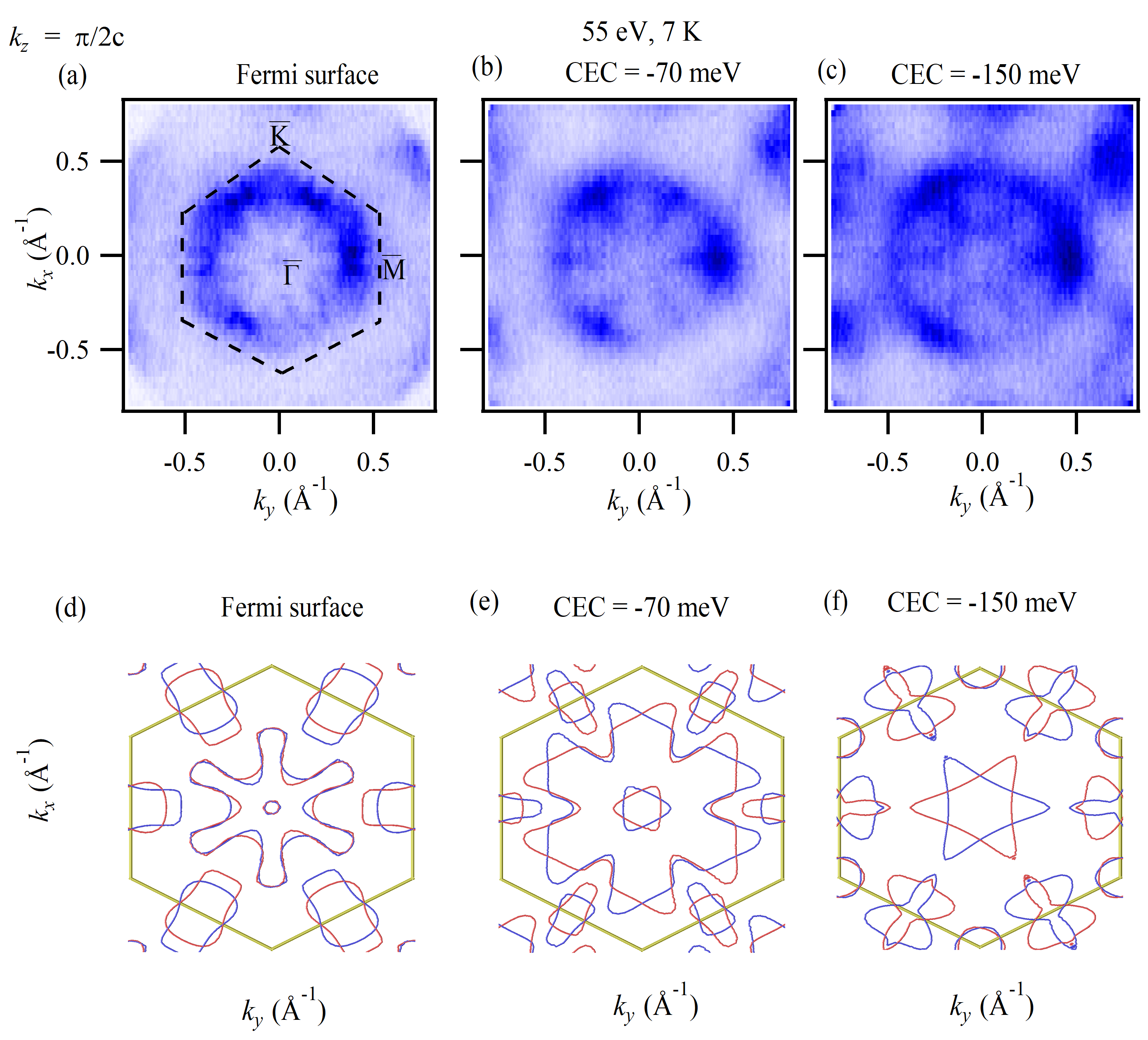} 
    \vspace{-1ex}
	\caption{Fermi surface and constant energy contours. (a-c) ARPES measured Fermi surface map and its constant energy contours at binding energies of - 70 meV and -150 meV, measured at a temperature of 7 K. (d-f) DFT calculated Fermi surface and constant energy contours, aligned with the experimental Fermi surface maps, calculated at k$_z$ = $\pi$/2c.}
\label{fig3}
\end{figure*}

Consequently, the Brillouin zone of CoNb$_4$Se$_8$ is half the size of that of NbSe$_2$ in the \textit{k$_x$-k$_y$} plane (see supplementary information for further details). Notably, the Co intercalation reduces the \textit{c}-axis lattice constant compared to the NbSe$_2$ unit cell, which correspondingly elongates the Brillouin zone (BZ) along this direction. Figure 1(c) displays the bulk BZ of CoNb$_4$Se$_8$ along with its projection onto the (001) surface, highlighting the relevant high-symmetry points. Figure 1(d) shows the core-level photoemission spectrum of cleaved CoNb$_4$Se$_8$, used for the ARPES experiments, displaying distinct peaks corresponding to Co, Nb, and Se. Figure 1(e) illustrates the variation of electrical resistivity, $\rho_{xx}$, with temperature for an electric current applied along the \textcolor{blue} {[100]}  direction. The resistivity exhibits a characteristic metallic trend, gradually decreasing from 300 K to 168 K. Below 168 K, it drops sharply, corresponding to the magnetic transition discussed in the next section, and eventually saturates at lower temperatures. The magnetic transition confirmed by single-crystal neutron diffraction in Ref \cite{GhimireCoNb4Se8} at 168 K is clearly visible in magnetic susceptibility measurements presented in Fig. 1(f). 

\begin{figure*}
	\includegraphics[width=15cm]{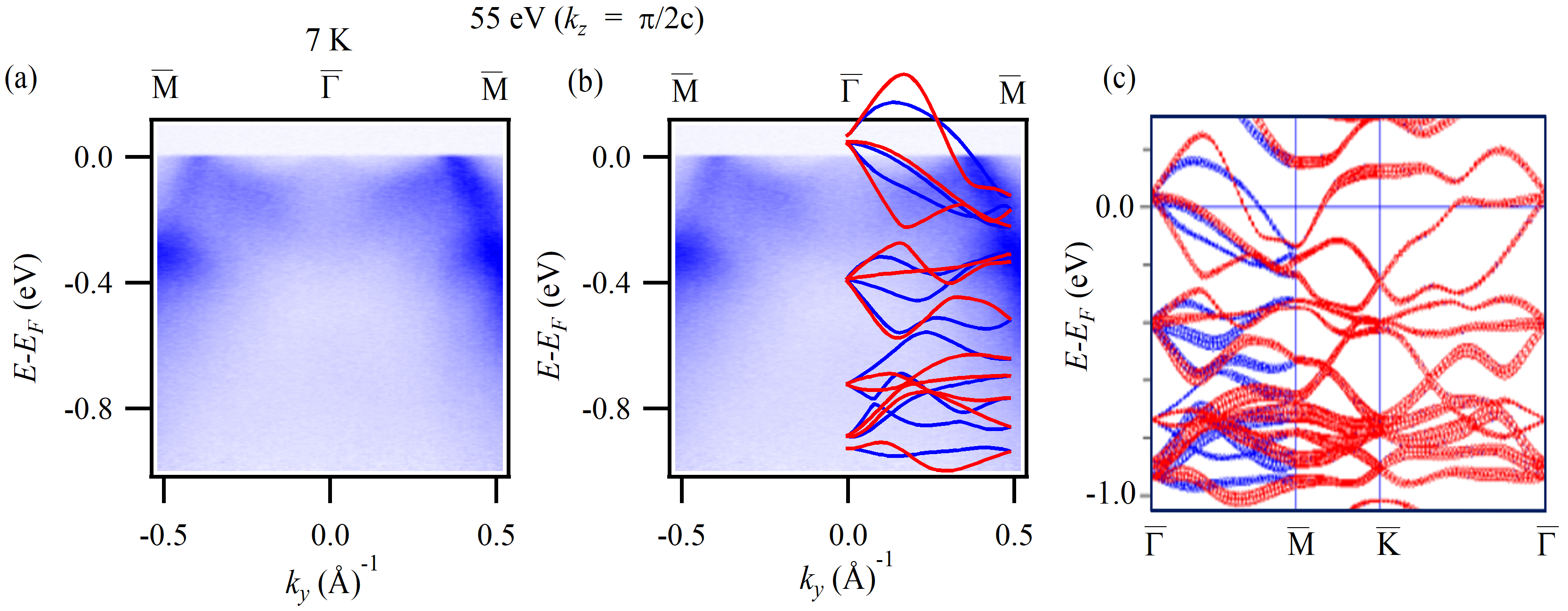} 
    \vspace{-1ex}
	\caption{Observation of altermagnetic band splitting. (a) Experimental band dispersion measured along the $\overline{\text{M}}-\overline{\Gamma}-\overline{\text{M}}$ high-symmetry line, measured at a photon energy of 55 eV, corresponding to the k$_z$ = $\pi$/2c plane at 7 K. (b) same as (a) but with the theoretical band structure overlaid, where red and blue curves indicate spin-up and spin-down bands, respectively. (c) Calculated band structure for the A-type AFM altermagnetic struture.}
\label{fig3}
\end{figure*} 

\indent To reveal the electronic structure of CoNb$_4$Se$_8$, we present ARPES measurements alongside DFT calculations, as shown in Figs. 2-4. We first examine the Fermi surface topology of CoNb$_4$Se$_8$. Figure 2(a), presents the large scale constant energy \textit{k$_y$-k$_z$} map, obtained with photon energies ranging from 40 eV to 110 eV along the $\overline{\text{M}}$--$\overline{\Gamma}$--$\overline{\text{M}}$ direction at a binding energy of -170 meV, using an inner potential of 15 eV. Clear periodic patterns are observed, allowing us to identify the positions of high-symmetry points, as marked by the labels ($\Gamma$ and A). Figure 2(b) shows the ARPES intensity mapping at E$_F$ as a function of \textit{k$_x$} and \textit{k$_y$}, measured with a photon energy of 50 eV. The dotted black line in Fig. 2(b) outlines the first BZ of CoNb$_4$Se$_8$ (see Supplementary Information for details). The Fermi surface reveals a small circular pocket centered at the $\overline{\Gamma}$ high-symmetry point. Additionally, dog-bone like Fermi pocket is observed around the $\overline{\text{M}}$ high-symmetry point. The previously reported DFT-calculated band structure revealed altermagnetic spin splitting along the $\overline{\Gamma}$--$\overline{\text{M}}$ direction \cite{GhimireCoNb4Se8}. The $\overline{\Gamma}$--$\overline{\text{M}}$ direction direction is along the Co-Co bonds. To investigate this, we focus our band dispersion measurements along the $\overline{\text{M}}$--$\overline{\Gamma}$--$\overline{\text{M}}$ and $\overline{\text{K}}$--$\overline{\Gamma}$--$\overline{\text{K}}$ high-symmetry lines as depicted in Fig. 2(c) and 2(d), respectively. The spectra display prominent dispersive features within an energy range of 0 to -0.7 eV along $\overline{\Gamma}$--$\overline{\text{M}}$ and 0 to -0.6 eV along $\overline{\Gamma}$--$\overline{\text{K}}$. The bands along both high-symmetry directions are steep and appear to cross the Fermi level. At the $\overline{\Gamma}$ point, a distinct W-shaped band is clearly visible in Figs. 2(c) and 2(d). Along the $\overline{\Gamma}$--$\overline{\text{K}}$ direction, two hole-like bands are observed: the outer band appears more diffuse, while the inner band is sharper, as indicated by the black and brown arrows, respectively. Both bands exhibit hole-like curvature, with the outer band remaining visible from E$_F$ to approximately -100 meV before fading.\\
\indent Next, we present the Fermi surface measurements at \textit{k$_z$} = $\pi$/2c, taken at temperature of 7 K, as shown in Fig. 3(a-c). For comparison, the Fermi surface calculated using DFT for \textit{k$_z$} = $\pi$/2c is also displayed, revealing altermagnetic splitting along the $\overline{\Gamma}$--$\overline{\text{M}}$ high-symmetry line, with spin-up and spin-down bands indicated by red and blue lines, respectively. The DFT-calculated Fermi surface shows good agreement with the experimental Fermi surface in Fig. 3(a) (see Supplementary Information for an overlay of the DFT Fermi surface and constant energy contours at \textit{k$_z$} = $\pi$/2c). Comparing the Fermi surfaces at 7 K with the DFT calculations, we observe the presence of dog-bone-shaped Fermi pockets at the $\overline{\text{M}}$ high-symmetry point, consistent with the DFT calculations shown in Fig. 3(d). To better illustrate the evolution of the Fermi surface maps at higher binding energies, we present constant energy contours at binding energies of -70 meV and -150 meV in Fig. 3(b) and 3(c), respectively. The central circular pocket at the $\overline{\Gamma}$ point expands with increasing binding energy, indicating the hole-like curvature of the band. The calculated constant energy contours in Fig. 3(e,f) qualitatively agree with the ARPES data shown in Fig. 3(b-c). The spin-split bands, represented by the red and blue curves, are clearly visible in the DFT calculations (Fig. 3a-c). \\ 

\indent Based on the photon-energy-dependent measurements (Fig. 2a), we plot the band dispersion cut corresponding to an intermediate \textit{k$_z$} value, where band splitting due to altermagnetic spin splitting is predicted to be most prominent. Figure 4(a-b) presents the band dispersions along the $\overline{\text{M}}$--$\overline{\Gamma}$--$\overline{\text{M}}$ high-symmetry line, measured using 55 eV photons, corresponding to the \textit{k$_z$} = $\pi$/2c plane. This measurement was performed at 7 K, within the altermagnetic regime. In Fig. 4(b), we overlay the DFT-calculated band dispersions on top of the ARPES data, focusing on the region where the band splitting is most pronounced. The ARPES data agrees well with the DFT-calculated bands, considering that the influence of photoemission matrix elements is not included in the DFT calculations. A direct comparison between the DFT calculations and the ARPES-measured band dispersions reveals clear band splitting along the $\overline{\text{M}}$--$\overline{\Gamma}$--$\overline{\text{M}}$ high-symmetry line. While the ARPES bands appear broader, the overall trend agrees with the theoretical predictions. For a detailed analysis of the band structure along various high-symmetry directions, the DFT-calculated band structure is presented in Fig. 4(c), where the altermagnetic spin-split bands are clearly visible. Further analysis using momentum distribution curves (see Supplementary Information) supports the presence of spin-split bands along this high-symmetry path, consistent with the predicted spin-split bands in the altermagnetic candidate CoNb$_4$Se$_8$.

\section{IV. Conclusions}
\indent In summary, we have conducted a comprehensive analysis of the electronic structure of the altermagnetic CoNb$_4$Se$_8$, which undergoes ordering below 168 K. The ARPES measured Fermi surface and band-dispersions are in good-agreement with our theoretical predictions. We provide evidence for the presence of altermagnetic spin-plitting of the bands using both DFT based calculations as well as our photon energy dependent ARPES measurements along the $\overline{\text{M}}$--$\overline{\Gamma}$--$\overline{\text{M}}$ high-symmetry line.  By demonstrating the band splitting, a crucial and important aspect of an altermagnet, this work provides a valuable platform for further exploration and understanding of altermagnetic materials.\\
\indent \textit{Note added}. During the preparation of this manuscript, we noticed that related studies on this material have also been posted on arXiv \cite{Dale, Vita, Candelora}.\\ 

\noindent \textbf{ACKNOWLEDGMENTS}\\
\indent Work performed by M.N., M.I.M., A.K.K., and H.S. was supported by the US Department of Energy (DOE), Office of Science, Basic Energy Sciences (BES) under Award DE-SC0024304. A.P.S., and M.S. acknowledge the support from the Air Force Office of Scientific Research MURI, Grant No. FA9550-20-1-0322. N.J.G., R.B.R., and I.I.M. were supported by Army Research Office under Cooperative Agreement Number W911NF- 22-2-0173. We thank Makoto Hashimoto and Donghui Lu for the beamline assistance at SSRL end station 5-2. The use of Stanford Synchrotron Radiation Lightsource (SSRL) in SLAC National Accelerator Laboratory is supported by the US Department of Energy, Office of Science, Office of Basic Energy Sciences under Contract No. DE-AC02-76SF00515.\\

\newpage
\def\bibsection{\section*{\refname}}

\vspace{2ex}


\end{document}